\documentclass{elsart}
\usepackage{graphicx}
\usepackage{amssymb}
\usepackage{amsmath}

\begin{document}

\setlength{\abovedisplayskip}{0.83\abovedisplayskip}
\setlength{\belowdisplayskip}{0.83\belowdisplayskip}
\setlength{\baselineskip}{0.96\baselineskip}

\begin{frontmatter}

\title{Recursive Structure and Bandwidth of Hales-Numbered Hypercube}
\author{Xiaohan Wang},
\ead{wangx28@mcmaster.ca}
\author{Xiaolin Wu},
\ead{xwu@ece.mcmaster.ca}
\address{Department of Electrical and Computer Engineering \\
McMaster University \\
Hamilton, Ontario, Canada, L8S 4K1}
\begin{abstract}
The Hales numbered $n$-dimensional hypercube and the corresponding
adjacency matrix exhibit interesting recursive structures in $n$.
These structures lead to a very simple proof of the well-known
bandwidth formula for hypercube, whose proof was thought to be
surprisingly difficult.  A related problem called hypercube
antibandwidth, for which Harper proposed an algorithm, is also
reexamined in the light of the above recursive structures, and a
close form solution is found.
\end{abstract}
\end{frontmatter}

\section{Introduction}

The problem of graph bandwidth has been extensively studied
\cite{survey,survey2}, and has found many applications such as
parallel computations, VLSI circuit design, etc.\  In this paper we
are particularly interested in the bandwidth of hypercubes. The
study of hypercube bandwidth can guide the design of communication
codes for error resilient transmission of signals over lossy
networks such as the Internet \cite{ISIT06}.

First, we restate the definitions of vertex numbering and graph
bandwidth, most of which are adopted from \cite{harper_book}.

\begin{defn}
A \emph{numbering} of a vertex set $V$ is any function
\begin{equation}
\eta : V \rightarrow \{1,2,\cdots,|V|\},
\end{equation}
which is one-to-one (and therefore onto).
\end{defn}
A numbering $\eta$ uniquely determines a total order, $\leq_\eta$,
on $V$ as: $u \leq_\eta v$ if $\eta(u) < \eta(v)$.  Conversely, a
total order defined on $V$ uniquely determines a numbering of the
graph.

\begin{defn}
The \emph{bandwidth of a numbering} $\eta$ of a graph $G=(V,E)$ is
\begin{equation}
bw(\eta) = \max_{\substack{\{u,v\} \in E }} |\eta(u) - \eta(v)|.
\end{equation}
\end{defn}

\begin{defn}
The \emph{bandwidth of a graph} $G$ is the minimum bandwidth over
all numberings, $\eta$, of $G$, i.e.
\begin{equation}
bw(G) = \min_{\eta} bw(\eta).
\end{equation}
\end{defn}
The graph of the n-dimensional cube, $Q^{(n)}$, has vertex set
$\{0,1\}^n$, the $n$-fold Cartesian product of $\{0,1\}$.  Thus
$|V_Q^{(n)}| = 2^n$.  $Q^{(n)}$ has an edge between two vertices
($n$-tuples of 0s and 1s) if they differ in exactly one entry.

\begin{defn}\label{def:hales}
The \emph{Hales order}, $\leq_H$, on $V_{Q^{(n)}}$, is defined by $u
\leq_H v$ if
\begin{enumerate}
  \item $w(u) < w(v)$, or
  \item $w(u) = w(v)$ and $u$ is greater than $v$ in lexicographic
  order relative to the right-to-left order of the coordinates,
\end{enumerate}
where $w(\cdot)$ is the Hamming weight of a vertex of $Q^{(n)}$.
This total order determines a numbering, $H^{(n)}:V_{Q^{(n)}}
\rightarrow \{1,2,\cdots,2^n\} $, which is called \emph{Hales
numbering}.
\end{defn}

\begin{thm}[Harper, \cite{harper_book}]\label{Thm:thm1}
The Hales numbering minimizes the bandwidth of the $n$-cube, i.e.
\begin{equation}\label{thm1}
bw(H^{(n)}) = bw(Q^{(n)}).
\end{equation}
\end{thm}
\begin{pf}
See Corollary 4.3 in \cite{harper_book}. \qed
\end{pf}

\begin{thm}[Harper, \cite{harper_book}]\label{cor1}
For the $n$-cube $Q^{(n)}$, we have
\begin{equation}\label{Eqn:cor1}
bw(Q^{(n)}) = \sum_{m=0}^{n-1} \binom{m}{\lfloor \frac{m}{2}
\rfloor}.
\end{equation}
\end{thm}

Although the above result has been known for forty years, no proof
seemed to appear in the literature.  Harper posed the proof of
Theorem~\ref{cor1} as an excise in his recent book
\cite{harper_book}, and noted ``it is surprisingly difficult".  In
the following section we present a rather simple proof.  The proof
also reveals some interesting effects of the Hales numbering on
hypercubes.

\section{Proof of the Bandwidth Formula for Hypercubes}

To prove Theorem~\ref{cor1}, we first need a lemma and some
definitions.

\begin{lem}\label{lem1}
We define a $2^n \times n$ $(0,1)$-matrix $S^{(n)}$ as
\begin{equation}\label{Eqn:Hales}
S^{(n)}=
\begin{bmatrix}
A_0^{(n)}\\
A_1^{(n)}\\
\vdots \\
A_n^{(n)}
\end{bmatrix},
\end{equation}
where $A_k^{(n)}$, $k=0,1,\cdots,n$, is an $\binom{n}{k}\times n$
$(0,1)$-matrix satisfying the following recursive formula
\begin{equation}
\label{Eqn:recursion} A_k^{(n)} = \begin{bmatrix} A_{k-1}^{(n-1)} &
\mathbf{1}
\\ A_{k}^{(n-1)} & \mathbf{0}
\end{bmatrix},\ k = 1, 2, \cdots, n-1,
\end{equation}
where $\mathbf{0}$ and $\mathbf{1}$ are column vectors containing
only $0$s and $1$s respectively.
As the base case, we have $A_0^{(n)}
%
= \mathbf{0}^T$ and $A_n^{(n)} = \mathbf{1}^T$. Then the row vectors
of $S^{(n)}$, from top to bottom, are all vertices of $Q^{(n)}$ in
the increasing Hales order.
\end{lem}

\begin{pf}
From Definition~\ref{def:hales}, it is sufficient to show that the
row vectors of $A_k^{(n)},\ k = 0,1,\cdots,n,$ are all distinct
vectors with Hamming weight $k$, which are sorted, from top to
bottom, in the decreasing lexicographic order.


We prove by induction on $n$.  The above assertion is trivially true
for $n=1$.  Assume the assertion holds for $n-1 \geq 1$.  Now for
$n$, $A_0^{(n)}$ is a vector of Hamming weight $0$ and $A_n^{(n)}$ a
vector of Hamming weight $n$, so the assertion trivially holds.  For
$1 \leq k \leq n-1$, the first $\binom{n-1}{k-1}$ vectors of
$A_k^{(n)}$ are all distinct and have Hamming weight $k$ by the
induction assumption that all row vectors in $A_{k-1}^{(n-1)}$ are
distinct and have Hamming weight $k-1$.  Further, these vectors are
in the decreasing lexicographic order because they share the same
rightmost bit and all vectors in $A_{k}^{(n-1)}$ are sorted.  By the
same argument the next $\binom{n-1}{k}$ vectors of $A_k^{(n)}$ are
distinct, of Hamming weight $k$, and sorted in the decreasing
lexicographic order as well.  Combining the above facts and
(\ref{Eqn:recursion}) concludes that the row vectors of $A_k^{(n)}$
are distinct, of Hamming weight $k$, and in the decreasing
lexicographic order. \qed
\end{pf}

\begin{defn}\label{def:am}
Given a graph $G=(V,E)$, for two vertex subsets $V_1 \subseteq V$
and $V_2 \subseteq V$ numbered by numberings $\eta_1$ and $\eta_2$
respectively, the \emph{adjacency matrix} of $V_1$ and $V_2$ is a
$|V_1| \times |V_2|$ matrix $M$ such that for any $u \in V_1$ and $v
\in V_2$
\begin{equation}
M(\eta_1(u),\eta_2(v)) = \left\{
                        \begin{array}{ll}
                          1 & \hbox{if $\{u,v\}\in E$;} \\
                          0 & \hbox{otherwise.}
                        \end{array}
                      \right.
\end{equation}
\end{defn}


\begin{defn}
The \emph{bandwidth of an $s \times t$ matrix} $M$ is the maximum
absolute value of the difference between the row and column indices
of a nonzero element of that matrix, i.e.
\begin{equation}
bw(M) = \max_{1 \leq i \leq s,\ 1 \leq j \leq t} \{ |i-j| \ | M(i,j)
\neq 0 \}.
\end{equation}
\end{defn}

\begin{rem}\label{rem:adjacencymatrix}
The bandwidth of a numbering $\eta$ of a graph $G$ is equal to the
bandwidth of the adjacency matrix of $G$ numbered by $\eta$.
\end{rem}

The bandwidth of a square matrix is obviously the maximum Manhattan
distance from a nonzero element to the main diagonal of the matrix.

\begin{figure}[thb]
\centering
\includegraphics[width=2.8in]{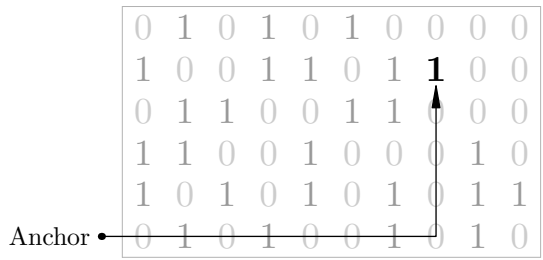}\\
\caption{Manhattan radius and anchor of a matrix.}\label{fig:anchor}
\end{figure}

\begin{defn}\label{def:metric}
For an $s\times t$ matrix $M$, its Manhattan radius $r(M)$ is
defined by
\begin{equation}\label{def:d}
r(M) = \max_{1 \leq i \leq s,\ 1 \leq j \leq t}\{ s-i+j\ |\
M(i,j)\neq 0\},
\end{equation}
which is the maximum Manhattan distance from a nonzero element of
$M$ to the position immediately to the left of the bottom-left
corner of matrix $M$ (an imaginary matrix element $M(s,0)$), as
shown in Fig.~\ref{fig:anchor}.  This imaginary matrix element
$M(s,0)$ is called the anchor of $M$.
\end{defn}

Let $M^{(n)}$ be the $2^n \times 2^n$ adjacency matrix of $Q^{(n)}$
numbered by $H^{(n)}$.  Recall from Lemma~\ref{lem1} that matrix
$S^{(n)}$ has as rows all vertices of $Q^{(n)}$ sorted by $H^{(n)}$.
Consider the submatrices $A^{(n)}_k$ and $A^{(n)}_{k'}$ in
$S^{(n)}$, and let the $\binom{n}{k}\times \binom{n}{k'}$ matrix
$M^{(n)}_{k,k'}$ be the adjacency matrix between $A^{(n)}_k$ and
$A^{(n)}_{k'}$.  Then $M^{(n)}_{k,k'}$, $0 \leq k, k' \leq n$, form
the $2^n \times 2^n$ adjacency matrix of the Hales numbered
hypercube: $M^{(n)}=[M^{(n)}_{k,k'}]$.  Obviously, $M^{(n)}_{k,k'}$
is an all-zero matrix if $|k-k'| \neq 1$.  Therefore, we have
\begin{equation}\label{Eqn:adjacency_mat}
M^{(n)} =
\begin{bmatrix}
\mathbf{0} & M^{(n)}_{0,1} & \mathbf{0} & \cdots & \mathbf{0} & \mathbf{0}\\
M^{(n)}_{1,0} & \mathbf{0} & M^{(n)}_{1,2} & \cdots & \mathbf{0} & \mathbf{0}\\
\mathbf{0} & M^{(n)}_{2,1} & \mathbf{0} & \cdots & \mathbf{0} & \mathbf{0}\\
\vdots & \vdots & \vdots & \ddots & \vdots & \vdots \\
\mathbf{0} & \mathbf{0} & \mathbf{0} & \cdots & \mathbf{0} & M^{(n)}_{n-1,n}\\
\mathbf{0} & \mathbf{0} & \mathbf{0} & \cdots & M^{(n)}_{n,n-1} & \mathbf{0}\\
\end{bmatrix}.
\end{equation}

The bandwidth of $M^{(n)}$ equals to the maximum Manhattan distance
from a nonzero element of $M^{(n)}$ to the main diagonal of
$M^{(n)}$.  Because of the symmetry of $M^{(n)}$, the bandwidth of
$M^{(n)}$ is equal to the maximum Manhattan distance from a nonzero
element of $M^{(n)}_{k,k+1}$, $k = 0,1,\cdots,n-1$, to the main
diagonal of $M^{(n)}$.  Note that the anchors of $M^{(n)}_{k,k+1}$
are all on the main diagonal.  Therefore, by
Definition~\ref{def:metric} the bandwidth of $M^{(n)}$ can be
expressed in terms of Manhattan radii of $M^{(n)}_{k,k+1}$:
\begin{equation}\label{Eqn:max}
bw(M^{(n)}) = \max_{k=0,\cdots,n-1} r(M^{(n)}_{k,k+1}).
\end{equation}

A pleasing recurrence structure of the Manhattan radius
$r(M^{(n)}_{k,k+1})$ affords us the following proof of
Theorem~\ref{cor1}.

\begin{pf}[Proof of Theorem~\ref{cor1}]
Because of Theorem~\ref{Thm:thm1} and
Remark~\ref{rem:adjacencymatrix}, we only need to show that the
bandwidth of the adjacency matrix of $Q^{(n)}$ with the Hales
numbering $H^{(n)}$ satisfies~(\ref{Eqn:cor1}).

Rewrite (\ref{Eqn:recursion}) as,
\begin{equation}
\label{Eqn:recursion1} A_k^{(n)} = \begin{bmatrix}
A_{k-1}^{(n-1)} & \mathbf{1} \\
A_{k}^{(n-1)} & \mathbf{0}
\end{bmatrix}\ \text{and}\
A_{k+1}^{(n)} = \begin{bmatrix} A_{k}^{(n-1)} & \mathbf{1}\\
A_{k+1}^{(n-1)} & \mathbf{0}
\end{bmatrix},\ k = 1, 2, \cdots, n-1.
\end{equation}
Then $M^{(n)}_{k,k+1}$, the adjacency matrix between $A_k^{(n)}$ and
$A_{k+1}^{(n)}$, can be divided into four submatrices.  The top-left
one is the adjacency matrix between $[A_{k-1}^{(n-1)}\ \mathbf{1}]$
and $[A_{k}^{(n-1)}\ \mathbf{1}]$, which equals to the adjacency
matrix between $A_{k-1}^{(n-1)}$ and $A_{k}^{(n-1)}$,
i.e.~$M^{(n-1)}_{k-1,k}$. Similarly, the bottom-right one is
$M^{(n-1)}_{k,k+1}$. Because there is no pair of Hamming distance
one between $A_{k-1}^{(n-1)}$ and $A_{k+1}^{(n-1)}$, the top-right
submatrix is an all-zero matrix. The bottom-left submatrix is the
adjacency matrix between $[A_{k}^{(n-1)}\ \mathbf{0}]$ and
$[A_{k}^{(n-1)}\ \mathbf{1}]$, which is an identity matrix
$\mathbf{I}_{\binom{n-1}{k}}$ of dimension $\binom{n-1}{k}$. Namely,
\begin{equation}\label{Eqn:recursion_matrix}
M^{(n)}_{k,k+1} =
\begin{bmatrix}
M^{(n-1)}_{k-1,k} & \mathbf{0} \\
\mathbf{I}_{\binom{n-1}{k}} & M^{(n-1)}_{k,k+1}\\
\end{bmatrix},\ k = 1,2,\cdots,n-1.
\end{equation}
Because $A_0^{(n)}$ is the all zero vector and $A_1^{(n)}$ contains
$n$ vectors of Hamming weight 1, we have $M^{(n)}_{0,1} =
\mathbf{1}^T$.

\begin{figure}[htb]
\centering
\includegraphics[width=3in]{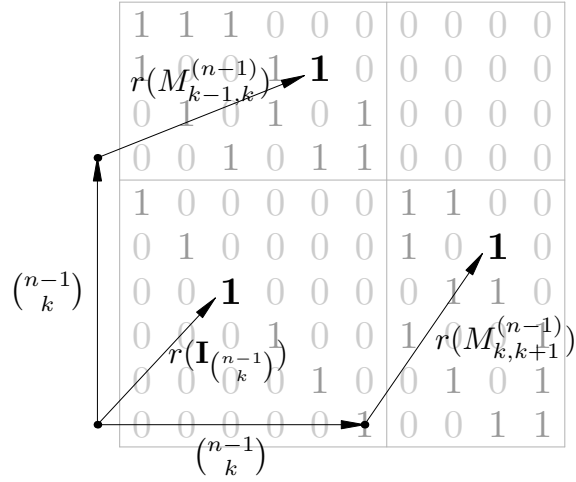}
\caption{The recursive structure of $r(M^{(n)}_{k,k+1})$, where $n =
5$ and $k = 2$.} \label{fig:M}
\end{figure}


It follows from (\ref{Eqn:recursion_matrix}) that the Manhattan
radius $r(M^{(n)}_{k,k+1})$ equals to the maximum Manhattan distance
from a nonzero element in submatrices $M^{(n-1)}_{k-1,k}$,
$M^{(n-1)}_{k,k+1}$ or $\mathbf{I}_{\binom{n-1}{k}}$ to the anchor
of $M^{(n)}_{k,k+1}$, as illustrated in Figure~\ref{fig:M}. From the
property of Manhattan distance and the fact that $r(\cdot)
> 0$, we have for $k = 1,2,\cdots,n-1$,
\begin{equation}\label{Eqn:3parts}
\begin{split}
r(M^{(n)}_{k,k+1}) &= \max\biggl\{ \binom{n-1}{k} +
r(M^{(n-1)}_{k-1,k}), \binom{n-1}{k} + r(M^{(n-1)}_{k,k+1}),
r(M^{(n-1)}_{k,k+1}) \biggr\} \\
&= \binom{n-1}{k} + \max\biggl\{r(M^{(n-1)}_{k-1,k}),
r(M^{(n-1)}_{k,k+1})\biggr\},
\end{split}
\end{equation}
and $r(M^{(n)}_{0,1}) = 1$ because $M^{(n)}_{0,1} = \mathbf{1}^T$.

Now we prove Theorem~\ref{cor1} by induction on $n$.  It is trivial
that when $n=1$, $r(M^{(1)}_{0,1}) = 1 = \binom{0}{0}$, and $ 0 =
\lfloor \frac{1-1}{2} \rfloor $. Assume that
\begin{equation}\label{Eqn:assume}
r(M^{(n-1)}_{k,k+1}) \leq \sum_{m=0}^{n-2} \binom{m}{\lfloor
\frac{m}{2} \rfloor},\ k=0,1,\cdots,n-2,
\end{equation}
where equality holds if $k = \lfloor \frac{n-1}{2} \rfloor$.  Then
we have
\begin{equation} \label{Eqn:induction}
\begin{split}
r(M^{(n)}_{k,k+1}) & = \binom{n-1}{k} +
\max\biggl\{r(M^{(n-1)}_{k-1,k}), r(M^{(n-1)}_{k,k+1})\biggr\} \\
& \leq \binom{n-1}{\lfloor \frac{n}{2} \rfloor} + \sum_{m=0}^{n-2}
\binom{m}{\lfloor \frac{m}{2} \rfloor} \\
& = \sum_{m=0}^{n-1} \binom{m}{\lfloor \frac{m}{2} \rfloor},
\end{split}
\end{equation}
in which the equality holds if $k = \lfloor \frac{n}{2} \rfloor$,
because when $n$ is even, $\lfloor \frac{n}{2} \rfloor - 1 = \lfloor
\frac{n-1}{2} \rfloor $, so $r(M^{(n-1)}_{k-1,k})$ achieves equality
in (\ref{Eqn:assume}); when $n$ is odd, $\lfloor \frac{n}{2} \rfloor
=\lfloor \frac{n-1}{2} \rfloor $, $r(M^{(n-1)}_{k,k+1})$ also
achieves equality in (\ref{Eqn:assume}). From (\ref{thm1}),
(\ref{Eqn:max}) and (\ref{Eqn:induction}), Theorem~\ref{cor1}
follows. \qed
\end{pf}

\section{Antibandwidth problem}

Another vertex numbering problem related to graph bandwidth is what
we call antibandwidth problem.  It is posed by reversing the
objective of vertex numbering in that we now want to maximize the
minimum distance between any adjacent pair of vertices.

\begin{defn}
The antibandwidth problem of a graph $G = (V,E)$ is defined as
\begin{equation}
f(G) = \max_{\eta} \min_{\{v,w\} \in E } |\eta(v) - \eta(w)|,
\end{equation}
where $\eta$ is a numbering of $G$.
\end{defn}

The antibandwidth problem has applications in code design for
communications \cite{ISIT06}.  On hypercubes the antibandwidth
problem has a very simple solution due to Harper \cite{harper}.

\begin{cor}[Harper, \cite{harper}]\label{lem2}
For the $n$-cube, first number the vertices with even Hamming
weights and then number the vertices with odd Hamming weights, in
the Hales order. The resulting numbering achieves $f(G)$.
\end{cor}
\begin{pf}
See \cite{harper}. \qed
\end{pf}

In this section, we provide a close form formula for the solution of
the antibandwidth problem on $n$-cubes, which is a new result.

\begin{thm}\label{cor2}
For the $n$-cube $Q^{(n)}$, we have
\begin{equation}
f(Q^{(n)}) = 2^{n-1} - \sum_{m=0}^{n-2} \binom{m}{\lfloor
\frac{m}{2} \rfloor}.
\end{equation}
\end{thm}

\begin{pf}
The numbering described in Corollary~\ref{lem2} determines a new
ordering of vertices
\begin{equation}\label{Eqn:Hales_max_min}
\tilde{S}^{(n)} =
\begin{bmatrix}
A_0^{(n)}\\
A_2^{(n)}\\
\vdots \\
A_{2\lfloor \frac{n}{2} \rfloor}^{(n)}\\
A_1^{(n)}\\
A_3^{(n)}\\
\vdots \\
A_{2\lfloor \frac{n-1}{2} \rfloor +1}^{(n)}\\
\end{bmatrix}.
\end{equation}
Similar to the proof of Theorem~\ref{cor1}, we have the adjacency
matrix of $Q^{(n)}$ with the vertices numbered in the order of
(\ref{Eqn:Hales_max_min})
\begin{equation}\label{Eqn:adjacency_mat2}
\tilde{M}^{(n)} =
\begin{bmatrix}
\mathbf{0} & \mathbf{0} & \cdots & \mathbf{0} & M^{(n)}_{0,1} & \mathbf{0} & \cdots & \mathbf{0}\\
\mathbf{0} & \mathbf{0} & \cdots & \mathbf{0} & M^{(n)}_{2,1} & M^{(n)}_{2,3 }& \cdots & \mathbf{0}\\
\vdots & \vdots & \ddots & \vdots & \vdots & \vdots & \ddots & \vdots\\
\mathbf{0} & \mathbf{0} & \cdots & \mathbf{0} & \mathbf{0} & \mathbf{0}& \cdots & M^{(n)}_{2\lfloor\frac{n}{2}\rfloor,2\lfloor\frac{n-1}{2}\rfloor+1}\\
M^{(n)}_{1,0} & M^{(n)}_{1,2} & \cdots & \mathbf{0} & \mathbf{0} & \mathbf{0} & \cdots & \mathbf{0}\\
\mathbf{0} & M^{(n)}_{3,2}& \cdots & \mathbf{0} & \mathbf{0} & \mathbf{0} & \cdots & \mathbf{0} \\
\vdots & \vdots & \vdots & \ddots & \vdots & \vdots & \ddots & \vdots\\
\mathbf{0} & \mathbf{0}& \cdots & M^{(n)}_{2\lfloor\frac{n-1}{2}\rfloor+1,2\lfloor\frac{n}{2}\rfloor} & \mathbf{0} & \mathbf{0} & \cdots & \mathbf{0}\\
\end{bmatrix}.
\end{equation}
From the symmetric structure of $\tilde{M}^{(n)}$, we only take into
account the lower part of the matrix $\tilde{M}^{(n)}$.  Then we
have
\begin{equation} \label{Enq:combinedelta}
f(\tilde{H}^{(n)}) = \min \biggl\{
\min_{k=1,3,\cdots,2\lfloor\frac{n-1}{2}\rfloor+1} \delta_{k,k-1},\
\min_{k=1,3,\cdots,2\lfloor\frac{n}{2}\rfloor-1}\delta_{k,k+1}
\biggr\},
\end{equation}
where $\delta_{k,k'}$ is the minimum Manhattan distance from a
nonzero element in the submatrix $M^{(n)}_{k,k'}$ to the main
diagonal of $\tilde{M}^{(n)}$.  Take into account a row of
submatrices $M^{(n)}_{k,\kappa},\ \kappa = k+1, k+3, \cdots, 2
\lfloor \frac{n}{2} \rfloor, 1, 3, \cdots, k$ and apply the property
of Manhattan distance, we have
\begin{equation} \label{Eqn:delta0}
\delta_{k,k+1} = \sum_{\kappa = k+1, k+3, \cdots, 2 \lfloor
\frac{n}{2} \rfloor, 1, 3, \cdots, k} W(M^{(n)}_{\kappa,k}) -
r(M^{(n)}_{k,k+1}),\ k=1,3,\cdots,2\lfloor\frac{n-1}{2}\rfloor+1
\end{equation}
where $W(\cdot)$ is the width of a matrix and hence
$W(M^{(n)}_{k,\kappa}) = \binom{n}{\kappa}$.  Therefore
\begin{equation} \label{Eqn:delta2}
\delta_{k,k+1} = \binom{n}{1}+ \binom{n}{3} + \cdots + \binom{n}{k}
+ \binom{n}{k+1}+\binom{n}{k+3}+\cdots+\binom{n}{2 \lfloor
\frac{n}{2} \rfloor } - r(M^{(n)}_{k,k+1}).
\end{equation}
Similarly, we have for $k=1,3,\cdots,2\lfloor\frac{n}{2}\rfloor-1$,
\begin{equation} \label{Eqn:delta1}
\delta_{k,k-1} = \binom{n}{1}+ \binom{n}{3} + \cdots + \binom{n}{k}
+ \binom{n}{k-1}+\binom{n}{k+1}+\cdots+\binom{n}{2\lfloor
\frac{n}{2} \rfloor} - r(M^{(n)}_{k,k-1}).
\end{equation}

From (\ref{Eqn:recursion}) and similar to the analysis of
(\ref{Eqn:recursion_matrix}), we derive the recursion form
\begin{equation}\label{Eqn:recursion_matrix2}
M^{(n)}_{k,k-1} =
\begin{bmatrix}
M^{(n-1)}_{k-1,k-2} & \mathbf{I}_{\binom{n-1}{k-1}} \\
\mathbf{0} & M^{(n-1)}_{k,k-1}.
\end{bmatrix},\ k = 2,\cdots,n,
\end{equation}
where as the base case $M^{(n)}_{1,0} = \mathbf{1}$.  The zero
matrix at the bottom-left corner has dimension $\binom{n-1}{k}
\times \binom{n-1}{k-2}$.  Therefore,
\begin{equation} \label{Eqn:k,k-1}
\begin{split}
r(M^{(n)}_{k,k-1}) &= \binom{n-1}{k} + \binom{n-1}{k-2} +
r(\mathbf{I}_{\binom{n-1}{k-1}}) \\
&= \binom{n-1}{k} + \binom{n-1}{k-2}+ \binom{n-1}{k-1}.
\end{split}
\end{equation}
Substituting $r(M^{(n)}_{k,k-1})$ of (\ref{Eqn:k,k-1}) into
(\ref{Eqn:delta1}), we have
\begin{equation} \label{Eqn:delta11}
\begin{split}
\delta_{k,k-1}  = & \binom{n}{1}+ \binom{n}{3} + \cdots +
\binom{n}{k} + \binom{n}{k-1}+\binom{n}{k+1}+\cdots+\binom{n}{2
\lfloor \frac{n}{2} \rfloor } \\
&- \binom{n-1}{k} - \binom{n-1}{k-2} - \binom{n-1}{k-1}  \\
= & 2^{n-1}, \\
\end{split}
\end{equation}
which can be easily established by considering the parity of $n$ and
using the binomial coefficients relations $\binom{n}{k} =
\binom{n-1}{k-1} +\binom{n-1}{k}$ and $\sum_{k=0,\cdots,n-1}
\binom{n-1}{k} = 2^{n-1}$.

Using the same justification and substituting $r(M^{(n)}_{k,k+1})$
of (\ref{Eqn:induction}) into (\ref{Eqn:delta2}), we have
\begin{equation} \label{Eqn:delta22}
\begin{split}
\delta_{k,k+1}  = & \binom{n}{1}+ \binom{n}{3} + \cdots +
\binom{n}{k} + \binom{n}{k+1}+\binom{n}{k+3}+\cdots+\binom{n}{2
\lfloor \frac{n}{2} \rfloor } \\
& - \binom{n-1}{k} -
\max\biggl\{r(M^{(n-1)}_{k-1,k}), r(M^{(n-1)}_{k,k+1})\biggr\}\\
= & 2^{n-1} - \max\biggl\{r(M^{(n-1)}_{k-1,k}), r(M^{(n-1)}_{k,k+1})\biggr\}\\
\geq & 2^{n-1} - \sum_{m=0}^{n-2} \binom{m}{\lfloor \frac{m}{2}
\rfloor},
\end{split}
\end{equation}
where equality holds when $k=\lfloor \frac{n-1}{2}\rfloor +1 $ or
$\lfloor \frac{n-1}{2}\rfloor$, whichever being odd.

Combining (\ref{Enq:combinedelta}), (\ref{Eqn:delta11}) and
(\ref{Eqn:delta22}) completes the proof. \qed
\end{pf}

\bibliographystyle{elsart-num}
\bibliography{harper}

\end{document}